\documentclass[aps,prl,twocolumn,balancelastpage,showpacs,reprint]{revtex4-1}
\usepackage{blindtext}
\usepackage{centernot}
\usepackage{graphicx}
\usepackage{amsmath}
\usepackage{times}
\usepackage{amssymb}
\usepackage{mathrsfs}
\usepackage{chemarr}
\usepackage{color}
\usepackage{url}
\usepackage{version}

\usepackage{mwe,tikz}\usepackage[percent]{overpic}

\usepackage{bm}
\usepackage[export]{adjustbox}
\definecolor{linkcolor}{rgb}{0,0,0.6}

\begin{document}

%%%%%%%%%%%%%%%%%%%%%%%%%%

\title{Field-Embedded Particles Driven by Active Flips}

\author{Ruben Zakine, Jean-Baptiste Fournier and Fr\'ed\'eric van Wijland}

\affiliation{Universit\'e Paris Diderot, USPC,
Laboratoire Mati\`ere et Syst\`emes Complexes (MSC), UMR 7057 CNRS, F-75205 Paris, France}

\date{\today}

\pacs{...}

\begin{abstract}
Systems of independent active particles embedded into a fluctuating environment are relevant to many areas of soft-matter science. We use a minimal model of noninteracting spin-carrying Brownian particles in a Gaussian field and show that activity-driven spin dynamics leads to patterned order. We find that the competition between mediated interactions and active noise alone can yield such diverse behaviors as phase transitions and microphase separation, from lamellar up to hexagonal ordering of clusters of opposite magnetization. These rest on complex multibody interactions. We find regimes of stationary patterns, but also dynamical regimes of relentless birth and growth of lumps of magnetization opposite to the surrounding one. Our approach combines Monte-Carlo simulations with analytical methods based on dynamical density functional approaches. 
\end{abstract}

\maketitle 

\begin{figure}
\includegraphics[width=.9\columnwidth]{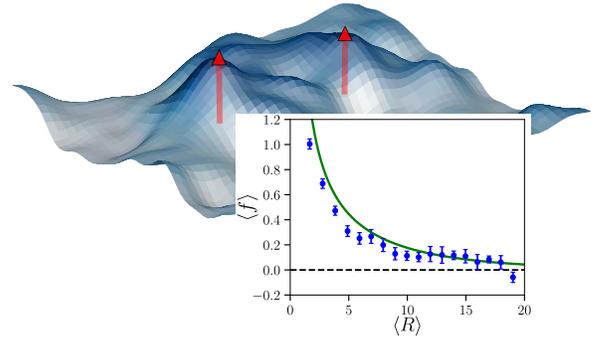}
\caption{(color online) Two particles (up spins) coupled to a fluctuating field (surface plot), favoring some value $\phi_0$ of the field. Inset: equilibrium average 
force as a function of the average particle separation, for the Hamiltonian described in the text. One particle is held fixed and the other one is tethered to a strong harmonic trap. Symbols: results of the numeric simulation (with incertitude). Solid line: analytical force deriving from $U(R)$ in the text. Parameters of the model: $r=0.01$, $\phi_0=8$, $B=1$ and $\mu=0.05$.}
\label{fig:forceBetweenInclusions}
\end{figure}

Active matter encompasses a broad class of physical systems, ranging from animal flocks~\cite{parrish97,Vicsek95,ballerini2008interaction,Chate08}, artificial self-propelled particles~\cite{Deseigne10,PhysRevLett.108.268303} and bacteria~\cite{Cates15} to molecular motors~\cite{Alvarado13}, and pumping~\cite{Prost96,Ramaswamy00} or multi-state particles such as proteins~\cite{Chen04}. While the former share the ability to extract energy from their environment and to convert it into directed motion, the latter can change conformation and exert active forces upon their surrounding medium (actin filaments, cell membrane).
Particles that deform a correlated elastic medium experience field-mediated interactions with a  fluctuation-induced component~\cite{Casimir48,Kardar99}, as illustrated in Fig.~\ref{fig:forceBetweenInclusions}. Mediated interactions occur for instance between interfaces, colloids or proteins in soft-matter media such as critical binary mixtures~\cite{Fisher78,Hertlein08}, liquid crystals~\cite{Ajdari91,Poulin97}, capillary interfaces~\cite{Nicolson49,Hu05Nat} and bio-membranes~\cite{Goulian93EPL,Dan93Lang,Dommersnes99EPJB,BitbolPlos12,vanderWel16}, including in nonequilibrium settings~\cite{lu2015out}.

An early approach to the question of why and how active particles, e.g., proteins in cell membranes, self-organize appeared in  \cite{Chen04,Chen06,Chen10}. In a parallel series of works on reactive two-state particle systems, spinodal decomposition coupled to active flips between the states has been shown to lead to a wealth of complex patterns. These have been described in~\cite{Oono87,Oono88}, \cite{Puri98}, \cite{Glotzer94,Glotzer95}, and \cite{Krishnan15}. A common feature to these approaches, necessary for the active flips to produce nontrivial patterns, is the requirement to start from directly interacting objects, either by assuming two-body interactions, or in a coarse-grained form by describing these in terms of an {\it ad hoc} Cahn-Hilliard field.

In this Letter we show that the emergence of activity-driven patterns can arise from purely field-mediated interactions, in the absence of any direct interactions between the particles. The nature of the coupling between the particles and the field is essential, as the existence of nonequilibrium phase transitions completely rests  on the physics governing the coupling. Furthermore, out of equilibrium, the coupling of particles to a field cannot be interpreted as effective direct interactions between particles. We concentrate on systems whose only nonequilibrium character resides in the active switching of the particles between two states coupling differently to the medium's field. We do not rely on a simplified field--particle density coupling. 
 We treat the field-particles interactions at the microscopic level (background illustration in Fig.~\ref{fig:forceBetweenInclusions}), in order to capture multi-body contributions and Casimir-like effects. 

In order to investigate such phenomena, we have striven to build up a model relying on the minimal necessary ingredients: Two populations of independent diffusing Ising particles, actively switching between their two states and interacting (quadratically) with a background Gaussian field, make up our model system. We refer to these particles as ASFIPs, {\it i.e.} active switching field interacting particles. The complexity of this system rests on the active nature of the particles, but also on the dynamics of the field-mediated interactions. We treat the dynamics of each particle and that of the field by equilibrium Langevin equations. The questions we ask are: i) Are induced interactions coupled to activity sufficient to generate emerging cooperative phenomena? ii) What is the role of activity (in as much as it drives us away from equilibrium) in generating complex patterns? iii) What is the role of multibody interactions and of Casimir-like forces in the states of matter that we observe?

%%%

We consider $N$ non-interacting particles at positions $\bm r_k(t)$, $1\leq k\leq N$, embedded in a medium whose elasticity is described by a scalar Gaussian field $\phi(\bm x,t)$. Our field $\phi$ might refer to a biomembrane thickness~\cite{Dan93Lang,BitbolPlos12} or internal lipid composition~\cite{KOMURA201434}. It may also refer to the shape of a biomembrane~\cite{Goulian93EPL,Dommersnes99EPJB} or to that of an interface under gravity~\cite{Nicolson49,Hu05Nat}. While all these systems are well  described by Gaussian fluctuating fields, the specifics of the Hamiltonian is model dependent. We choose the simplest model, with energy
\begin{align}
H_0=\int\! d^2 x \left[\frac r 2 \phi^2+ \frac c 2 (\bm \nabla\phi)^2\right].
\end{align}
To model particles that can be in two states, we attach a spin variable $S_k=\pm1$ to each particle. The underlying picture we have in mind is that of protein inclusions changing conformation through external chemical activity~\cite{allostery}. The particle-field coupling is a key-ingredient, we take:
\begin{align}
H_\mathrm{int}=\sum_{k=1}^N \frac B 2 \left(\phi(\bm r_k)- S_k\phi_0\right)^2.
\label{eq_hamiltonian}
\end{align}
The effect of this interaction is to adjust locally the field to a spin-dependent amplitude $\pm\phi_0$, with a strength governed by the stiffness coefficient $B$. We draw the reader's attention to the quadratic nature of $H_\mathrm{int}$. Linear couplings in the field are quite unrealistic as they miss multibody and fluctuation-induced interactions. We do not wish to discard such ingredients that exist in real systems. The total energy becomes $H=H_0+H_\mathrm{int}$. We purposely omit excluded volume or any other kind of direct interaction, which allows us to witness field-induced phenomena only.

We endow $\phi$ with a purely relaxational dynamics satisfying detailed balance:
\begin{align}
&\partial_t \phi(\bm x,t)=-\Gamma\frac{\delta H}{\delta \phi(\bm x,t)}+\sqrt{2\Gamma T}\,\xi(\bm x,t),\label{eq_field_evol}
\end{align}
where $T$ is the temperature in energy units, $\Gamma$ the field mobility and $\xi(\bm x,t)$ a Gaussian white noise. Particles diffuse according to equilibrium overdamped Langevin equations:
\begin{align}
&\frac{d\bm r_k}{d t}=-\mu \frac{\partial H}{\partial \bm r_k}+\sqrt{2\mu T}\bm\eta_k(t),\label{eq_langevin}
\end{align}
where $\mu$ is a mobility coefficient (assumed to be spin and field independent), and the $\bm \eta_k(t)$'s are independent Gaussian white noises. We use the simplifying assumption that $\bm \eta_k$ and $\xi$ are independent (as is generic in soft matter, see, {\it e.g.} Ref.~\cite{Ramaswamy00} for proteins in biomembranes). 

Finally, the out-of-equilibrium dynamics arises from the internal degree of freedom of the particles. Each particle flips through the action of an external energy source (e.g., photons, chemical reactions), with fixed rates:
\begin{align}\label{spin-flips}
S_k=\mathrm{-1}\, \xrightleftharpoons[~\gamma~]{~\alpha~} \,S_k=\mathrm{+1}.\qquad\text{(ASFIP)}
\end{align}
This is the one process breaking detailed balance for ASFIPs due to the coupling with the field and particle dynamics. 

Since we want to understand how our system behaves exactly, taking into account detailed out-of-equilibrium mediated interactions, multibody and fluctuation-induced effects without relying on approximate analytical methods, we first perform Monte Carlo simulations. We discretize our equations on a lattice with spacing $a$ with the normalization $a=T=\Gamma=c=1$ (see SM, Sec.~I). The remaining parameters are $r$, fixing the field's correlation length $r^{-1/2}$, $B$ the stiffness of the spin--field coupling, $\phi_0$ the targeted field, and the dynamical parameters $\mu$, $\alpha$ and $\gamma$, all scaled by the field's mobility.

We implement discrete time Monte Carlo simulations on a two dimensional (2D) square lattice of size $L\times L$ with periodic boundary conditions, as detailed in the Supplemental Material (SM, Sec.~II). The field is defined on the lattice sites and the particles move from site to adjacent site. Between times $t$ and $t+\Delta t$, particles can hop, or flip spin, or stay on the same site. To take into account the relative dynamics of the particles and the field, we implement a tower sampling algorithm~\cite{krauth2006statistical} instead of a Metropolis one.  

In order to characterize the field-mediated interaction in equilibrium, we first study the force exchanged by two particles a distance $R$ apart in the manner described in Fig.~\ref{fig:forceBetweenInclusions} (or Sec.~III in SM for a precise description). The effective potential $U(R)$ between these two particles can be derived (see SM) from a field-theoretic calculation. As shown in Fig.~\ref{fig:forceBetweenInclusions}, the force is well fitted by $U'(R)$, which confirms the validity of our Monte Carlo simulation.  The force is attractive for equal spins and decays typically over the field correlation length. We found that for $R \geq 1$ and $\phi_0\gtrsim 3$ the fluctuation-induced component of the force is negligible, but this does not mean that it must be so out of equilibrium. Actually, the standard deviation of the force, which has a component coming from the Langevin force on the particle and another coming from the fluctuations of the field, is much larger than its average. Note that whereas in equilibrium the field samples thermally all of its configurations (even when the particles move), in the out-of-equilibrium case the dynamics of the field could yield retarded effects with important consequences.
 
%%%%%

\begin{figure}
\includegraphics[width=\columnwidth]{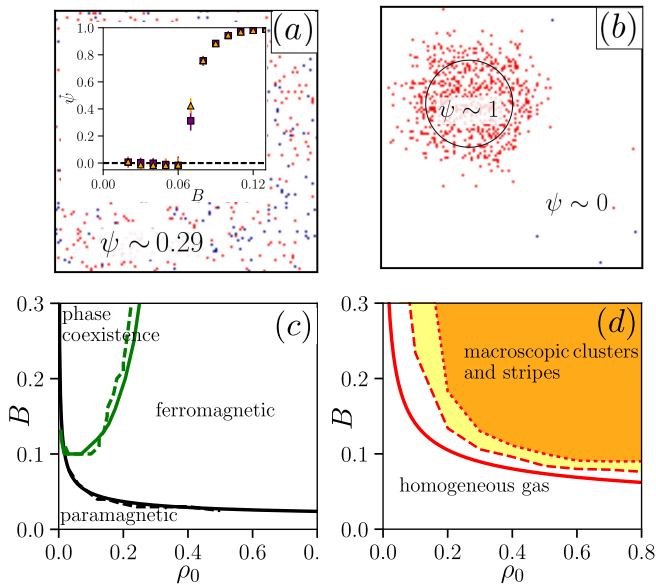}
\caption{(color online) (a) SFIPs in a box with periodic boundary conditions, $L=150$, $\rho_0=0.05$, $r=0.01$, $\phi_0=8$ and $B=0.07$, yielding a ferromagnetic state. Red (blue) dots indicate particles with up (down) spins. Inset: magnetization order parameter as a function of $B$. Light orange (dark purple) symbols correspond to $B$ increasing (decreasing). (b) SFIPs for $B=0.26$ (same other parameters) showing the coexistence of a ferromagnetic liquid and a paramagnetic gas (c) Phase diagram of SFIPs in terms of total density and coupling strength for $r=0.01$, $\phi_0=8$ and $\mu=5$. Solid lines: mean-field predictions for the paramagnetic--ferromagnetic transition (black) and for the binodal curve of the phase separation (green, or gray). The corresponding dashed lines are the results of the Monte Carlo simulations. (d) ASFIPs for the same parameters and $\alpha=\gamma=0.1$. Solid red line: mean-field prediction for the transition to a patterned phase. Yellow (light gray) zone: beginning of segregation. Orange (gray) zone: ferromagnetic stripes and macroscopic clusters.}
\label{fig:snapshotSfip}
\end{figure}

%%%%%%%%%%%

Before we embark into a full description of the out-of-equilibrium ASFIPs, we wish to introduce their equilibrium counterpart, for future comparison purposes. In equilibrium, Switching Field Interacting Particles (SFIPs) have  transition rates $\propto \exp{(\pm w_k)}$ with $w_k=B\phi_0\phi(\bm r_k)$ half the energy variation in a spin flip. Such particles experience equilibrium field mediated-interactions and flips, while they diffuse on the lattice. Let $N$ be the total number of particles and $\rho_0=N/L^2$. At fixed $r$ and $\phi_0$, we increase the coupling strength $B$. We observe first a paramagnetic--ferromagnetic phase transition (Fig.~\ref{fig:snapshotSfip}a), then a phase separation into a dense ferromagnetic fluid coexisting with a paramagnetic gas (Fig.~\ref{fig:snapshotSfip}b). These states obviously do not depend on the dynamical parameter $\mu$. We characterize the magnetization of each homogeneous phase by the order parameter $\psi=\langle\rho^+-\rho^-\rangle/\langle\rho\rangle$, where $\rho^\pm$ is the density of particles with $\pm1$ spins and $\rho=\rho^++\rho^-$, and we find that the paramagnetic--ferromagnetic phase transition is compatible with a continuous one (Fig.~\ref{fig:snapshotSfip}a, inset).

%%%%%%%%%%%

Since SFIPs are in equilibrium, we can rely on thermodynamics to study their behavior.
The mean-field energy density naturally deriving from $H$ is 
\begin{align}
f_\mathrm{mf}&=\frac r 2 \phi^2+\frac B 2 \rho^+(\phi-\phi_0)^2+\frac B 2 (\rho-\rho^+)(\phi+\phi_0)^2 \nonumber\\
&+\rho^+\ln \rho^+ +(\rho-\rho^+)\ln(\rho-\rho^+).
\end{align} 
Since $\rho$ is the only conserved quantity, we minimize $f_\mathrm{mf}$ with respect to $\phi$ and $\rho^+$, which yields an energy density $f'_\mathrm{mf}(\rho)$ and $\phi=B\phi_0(2\rho^+-\rho)/(r+B\rho)$ with either $\rho^+=\rho^-=\rho/2$ (paramagnetic phase) or $\rho^+\ne\rho^-$ (ferromagnetic phase). At low values of $B$, the system is uniform and there is a continuous paramagnetic--ferromagnetic transition at $
B_c^{(\mathrm{mf})}=(1+\sqrt{1+4r\,\phi_0^2/\rho_0})/(2\phi_0^2)$.
At higher values of $B$, we obtain through the double tangent construction on $f'_\mathrm{mf}(\rho)$ a phase separation between a low density paramagnetic phase and a high density ferromagnetic phase. These mean-field predictions correspond to the continuous lines of
Fig.~\ref{fig:snapshotSfip}c while the results of the Monte Carlo simulations are indicated by the dashed lines. The agreement is all the better as we are working at large $\phi_0$ or low $T$ 
%%%

%%%%%%%%%We obtain the same results whether we simulate with an initial state  SFIPs and suddenly impose the rates $\alpha$ and $\gamma$ or we start with a homogeneous gas of ASFIPs.

\begin{figure}
\includegraphics[width=.9\columnwidth]{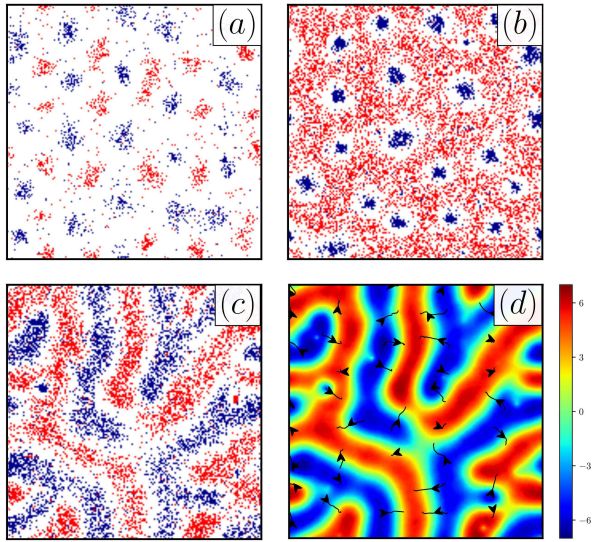}
\caption{(color online) Snapshots of the patterns created by ASFIPs. The parameters are $r=0.01$, $\phi_0=8$.  Red (blue) dots indicate particles with up (down) spins. (a) Square phase obtained for symmetric flips ($B=0.26$, $\rho_0=0.1$, $\mu=2.5$, $\gamma=\alpha=0.005$, $L=180$). (b) Hexagonal phase of clusters ($B=0.15$, $\rho_0=0.4$, $\mu=5.0$, $\alpha=0.02$, $\gamma=\alpha/3$, $L=160$). (c) Striped phase obtained for symmetric flips ($B=0.15$, $\rho_0=0.4$, $\mu=5.0$, $\gamma=\alpha=0.02$, $L=160$). (d) shows the corresponding $\phi$ field map of (c) and the time average of the fluxes of spin-up particles.}
\label{fig:asfip_snapshot}
\end{figure}

We now return to our original nonequilibrium ASFIPs. The phase diagram of ASFIPs undergoing symmetric flips ($\alpha=\gamma$) is shown in Fig.~\ref{fig:snapshotSfip}d. The system is always paramagnetic on global average, due to the imposed flips, however increasing $B$ at fixed $\rho_0$ yields first a transition from a paramagnetic gas to ferromagnetic clusters of either magnetizations, as illustrated in Fig.~\ref{fig:asfip_snapshot}a, then to a phase of dynamical ferromagnetic stripes.  A typical snapshot of the macroscopic stripes is shown in Fig.~\ref{fig:asfip_snapshot}c.  
For asymmetric flips (e.g., $\alpha=3\gamma$) we observe a dynamical hexagonal pattern of clusters  (Fig.~\ref{fig:asfip_snapshot}b). These clusters are formed by the particles with the higher flip rate. 

% Lower rates yield patterns of larger sizes, although given the limited size of our simulations we cannot say whether there is a transition to a macroscopic phase separation (as is the case in an equilibrium limit in which spins are not allowed to flip) or if the pattern size increases indefinitely. Conversely, if the flipping rates are too large, patterns are destroyed and the system becomes homogeneous.

To gain insight into the physics of this pattern creation, we have computed the average fluxes of the particles and the map of the $\phi$ field (Fig.~\ref{fig:asfip_snapshot}d). First, we see that high (low) field regions have a majority of spin up (spin down) particles. Hence, ASFIPs also tend to phase separate due to the field mediated interactions. We observe that spin up particles travel from regions of low spin-up density  to regions of high spin-up density, just as for the coarsening of the equilibrium SFIP's. However, these activity-driven fluxes never vanish, which is specific to being out of equilibrium. Therefore, whenever a particle flips, it is expelled by the field-mediated interactions towards the nearest region  matching its updated spin. This is the mechanism by which pattern formation occurs.

In addition, within large enough regions of a given magnetization, we observe the systematic nucleation and growth of lumps of opposite magnetization ({\it e.g.}, small visible blue islands in Fig.~\ref{fig:asfip_snapshot}b or red and blue ones in \ref{fig:asfip_snapshot}c), as illustrated by the movies in the SM, Sec.~V. They diffuse, get expelled and eventually merge into a domain of the same magnetization. The mechanism allowing for this behavior is intrinsically out of equilibrium. For SFIPs, energy balance quickly prevents the growth of lumps, whereas for ASFIPs long lived spin states are allowed to gather and form the seed for a dynamic lump which then grows by accretion.

%%%

What is the importance of fluctuation-induced interactions and multi-body effects in the ASFIP system? If we turn off the field noise $\xi(\bm x,t)$ (while keeping the dynamics on the particles unchanged), we observe that particle segregation and pattern formation occur as soon as $B$ exceeds the mean-field threshold (solid red line in Fig.~\ref{fig:snapshotSfip}d). Thermal fluctuations tend to destroy patterns and fluctuation-induced forces are too weak to play any pattern-favoring role. In order to investigate multi-body effects, we have also replaced the quadratic coupling of Eq.~\eqref{eq_hamiltonian} with a linear coupling adjusted to yield, up to a very good approximation, the same two-body field mediated interaction (see SM). This results in the condensation of the particles on a unique site for SFIP's and in the absence of activity driven patterns for ASFIP's. Multibody interactions are thus essential. We have checked that adding a hard-core repulsion, in the quadratic coupling case, has almost no effect on the phase diagram, indicating that modest short-range interaction are irrelevant in our system.

Let us rationalize our findings on the phase diagram with a dynamical mean-field approach. Since  ASFIPs diffuse by means of overdamped Langevin equations, we implement a 
Dean--Kawasaki~\cite{dean1996langevin,KAWASAKI199435} approach in the noiseless limit. The evolution equations then read $\partial \rho^\pm+\bm\nabla\cdot \bm j^\pm=0$ with $\bm j^\pm=-\mu \rho^\pm \bm\nabla(\partial f_\mathrm{mf}/\partial \rho^\pm)$. Taking spin exchange into account, we arrive at the evolution equations:
\begin{align}
\partial_t\rho^\pm&=
\mu \nabla^2\rho^\pm\nonumber+\mu B\,\bm\nabla\cdot\left[\rho^\pm(\phi\mp\phi_0)\bm\nabla \phi\right]\pm\alpha\rho^-\mp\gamma\rho^+,\nonumber
\\
\partial_t\phi&= \nabla^2 \phi - r\phi- B \rho^+(\phi-\phi_0)- B \rho^-(\phi+\phi_0).
\end{align}
Linear stability analysis (LSA) shows that above a threshold in $B$ the stationary and homogeneous solution [$\rho_s^+=\rho_0\alpha/(\alpha +\gamma)$, $\rho_s^-=\rho_0\gamma/(\alpha +\gamma)$ and $\phi_s=B\phi_0(\rho_s^+-\rho_s^-)/(r+B\rho_0)$] is no longer stable, indicating the onset of a patterned phase. For symmetric flips, $\gamma=\alpha$, this threshold reads $
B_a^{(\mathrm{mf})}=(1+\sqrt{1+4r\phi_0^2/\rho_0+4\phi_0 s}+2\phi_0s
)/(2\phi_0^2)$, where $s=\sqrt{2\alpha/(\mu\rho_0)}$. The agreement with the results of the Monte Carlo simulations is satisfying (Fig.~\ref{fig:snapshotSfip}d). 

In dimensionful form, $B_a^{(\mathrm{mf})}$ turns out to be independent of the field mobility $\Gamma$. We have checked this property in the Monte Carlo simulations and found indeed that varying $\Gamma$ over five orders of magnitude has no effect on the phase diagram. The times scales involved in the pattern formation, however, depend on $\Gamma$. In addition, LSA shows that varying dynamical parameters $\Gamma$, or $\mu$ and $\alpha$ while keeping $s$ constant, does not change the interval $[k_\mathrm{min},k_\mathrm{max}]$ over which temporal eigenvalues destabilize the homogeneous solution. We also found that increasing the particles' mobility $\mu$ (while keeping all other parameters fixed) enlarges the  domain where patterns are stable in the phase diagram.  We expect that taking the  $s\to0$ limit in $B_a^{(\mathrm{mf})}$ leads us to equilibrium. Indeed, when sending $\alpha\to 0$ or $\mu\to\infty$, the field is effectively sampled in an equilibrium manner. In between two flips, the diffusion of the particles is a quasi-equilibrium process. Furthermore, LSA confirms that patterns are specific to out-of-equilibrium since sending $s\to 0$ yields $k_\mathrm{min}\to 0$, and then we end up with a more conventional coarsening of the binary mixture in that regime.
What is more surprising, however, is that $B_a^{(\mathrm{mf})}\to B_c^{(\mathrm{mf})}$ when $s\to 0$. This is due to the specific choice $\alpha=\gamma$, which protects the up-down symmetry, hence leaving the phase boundary unchanged in the equilibrium limit (the scenario does not hold for $\alpha\neq\gamma$).

% Finally, we have also implemented effective pairwise forces between particles by the mean of a linear coupling between the particles and the field. Even though the force can be tuned to be quantitatively equivalent, a linear coupling lead to infinite condensation, underlying at the same time the importance of $N$-body effects along with induced fluctuations in our model system.

We are now in a position to summarize the  answers to our original questions. Starting from a microscopic model where {\it  noninteracting} particles are coupled to a Gaussian field, we have proved that field-mediated interactions combined with activity can generate a wealth of new emerging cooperative phenomena. This is relevant for soft-matter systems in which interactions are mostly indirect, field-mediated. In our system, it is the presence of activity which drives complex patterns of particle clusters by a continuous tossing in and out of diffusing particles. The quadratic coupling that we have used captures both multibody and fluctuation-induced interactions. While the former is of paramount relevance, the latter is entirely dominated by thermal fluctuations and can be neglected. We see several directions along which we could expand our findings. From a theoretical standpoint, we wish to investigate the effect of varying the details of the correlator (membranes will feature higher derivatives for instance). Similarly, the particle--field coupling may also involve higher derivatives depending on physical context. Exploring the consequences of hydrodynamic effects is also of great relevance. Finally, it would be interesting to investigate such emerging phenomena in experimental systems of active particles, even in athermal macroscopic systems where activity alone might suffice.

%\bibliographystyle{apsrev4-1}
%\bibliography{asfipnewBiblio}

\begin{thebibliography}{0}%
\makeatletter
\providecommand \@ifxundefined [1]{%
 \@ifx{#1\undefined}
}%
\providecommand \@ifnum [1]{%
 \ifnum #1\expandafter \@firstoftwo
 \else \expandafter \@secondoftwo
 \fi
}%
\providecommand \@ifx [1]{%
 \ifx #1\expandafter \@firstoftwo
 \else \expandafter \@secondoftwo
 \fi
}%
\providecommand \natexlab [1]{#1}%
\providecommand \enquote  [1]{``#1''}%
\providecommand \bibnamefont  [1]{#1}%
\providecommand \bibfnamefont [1]{#1}%
\providecommand \citenamefont [1]{#1}%
\providecommand \href@noop [0]{\@secondoftwo}%
\providecommand \href [0]{\begingroup \@sanitize@url \@href}%
\providecommand \@href[1]{\@@startlink{#1}\@@href}%
\providecommand \@@href[1]{\endgroup#1\@@endlink}%
\providecommand \@sanitize@url [0]{\catcode `\\12\catcode `\$12\catcode
  `\&12\catcode `\#12\catcode `\^12\catcode `\_12\catcode `\%12\relax}%
\providecommand \@@startlink[1]{}%
\providecommand \@@endlink[0]{}%
\providecommand \url  [0]{\begingroup\@sanitize@url \@url }%
\providecommand \@url [1]{\endgroup\@href {#1}{\urlprefix }}%
\providecommand \urlprefix  [0]{URL }%
\providecommand \Eprint [0]{\href }%
\providecommand \doibase [0]{http://dx.doi.org/}%
\providecommand \selectlanguage [0]{\@gobble}%
\providecommand \bibinfo  [0]{\@secondoftwo}%
\providecommand \bibfield  [0]{\@secondoftwo}%
\providecommand \translation [1]{[#1]}%
\providecommand \BibitemOpen [0]{}%
\providecommand \bibitemStop [0]{}%
\providecommand \bibitemNoStop [0]{.\EOS\space}%
\providecommand \EOS [0]{\spacefactor3000\relax}%
\providecommand \BibitemShut  [1]{\csname bibitem#1\endcsname}%
\let\auto@bib@innerbib\@empty
%</preamble>
\end{thebibliography}%


\begin{thebibliography}{0}
\expandafter\ifx\csname natexlab\endcsname\relax\def\natexlab#1{#1}\fi
\expandafter\ifx\csname bibnamefont\endcsname\relax
  \def\bibnamefont#1{#1}\fi
\expandafter\ifx\csname bibfnamefont\endcsname\relax
  \def\bibfnamefont#1{#1}\fi
\expandafter\ifx\csname citenamefont\endcsname\relax
  \def\citenamefont#1{#1}\fi
\expandafter\ifx\csname url\endcsname\relax
  \def\url#1{\texttt{#1}}\fi
\expandafter\ifx\csname urlprefix\endcsname\relax\def\urlprefix{URL }\fi
\providecommand{\bibinfo}[2]{#2}
\providecommand{\eprint}[2][]{\url{#2}}

\end{thebibliography}


\begin{thebibliography}{38}%
\makeatletter
\providecommand \@ifxundefined [1]{%
 \@ifx{#1\undefined}
}%
\providecommand \@ifnum [1]{%
 \ifnum #1\expandafter \@firstoftwo
 \else \expandafter \@secondoftwo
 \fi
}%
\providecommand \@ifx [1]{%
 \ifx #1\expandafter \@firstoftwo
 \else \expandafter \@secondoftwo
 \fi
}%
\providecommand \natexlab [1]{#1}%
\providecommand \enquote  [1]{``#1''}%
\providecommand \bibnamefont  [1]{#1}%
\providecommand \bibfnamefont [1]{#1}%
\providecommand \citenamefont [1]{#1}%
\providecommand \href@noop [0]{\@secondoftwo}%
\providecommand \href [0]{\begingroup \@sanitize@url \@href}%
\providecommand \@href[1]{\@@startlink{#1}\@@href}%
\providecommand \@@href[1]{\endgroup#1\@@endlink}%
\providecommand \@sanitize@url [0]{\catcode `\\12\catcode `\$12\catcode
  `\&12\catcode `\#12\catcode `\^12\catcode `\_12\catcode `\%12\relax}%
\providecommand \@@startlink[1]{}%
\providecommand \@@endlink[0]{}%
\providecommand \url  [0]{\begingroup\@sanitize@url \@url }%
\providecommand \@url [1]{\endgroup\@href {#1}{\urlprefix }}%
\providecommand \urlprefix  [0]{URL }%
\providecommand \Eprint [0]{\href }%
\providecommand \doibase [0]{http://dx.doi.org/}%
\providecommand \selectlanguage [0]{\@gobble}%
\providecommand \bibinfo  [0]{\@secondoftwo}%
\providecommand \bibfield  [0]{\@secondoftwo}%
\providecommand \translation [1]{[#1]}%
\providecommand \BibitemOpen [0]{}%
\providecommand \bibitemStop [0]{}%
\providecommand \bibitemNoStop [0]{.\EOS\space}%
\providecommand \EOS [0]{\spacefactor3000\relax}%
\providecommand \BibitemShut  [1]{\csname bibitem#1\endcsname}%
\let\auto@bib@innerbib\@empty
%</preamble>
\bibitem [{par(1997)}]{parrish97}%
  \BibitemOpen
  \href {\doibase 10.1017/CBO9780511601156} {\emph {\bibinfo {title} {Animal
  Groups in Three Dimensions: How Species Aggregate}}}\ (\bibinfo  {publisher}
  {Cambridge University Press},\ \bibinfo {year} {1997})\BibitemShut {NoStop}%
\bibitem [{\citenamefont {Vicsek}\ \emph {et~al.}(1995)\citenamefont {Vicsek}
  \emph {et~al.}}]{Vicsek95}%
  \BibitemOpen
  \bibfield  {author} {\bibinfo {author} {\bibfnamefont {T.}~\bibnamefont
  {Vicsek}} \emph {et~al.},\ }\href {\doibase 10.1103/PhysRevLett.75.1226}
  {\bibfield  {journal} {\bibinfo  {journal} {Phys. Rev. Lett.}\ }\textbf
  {\bibinfo {volume} {75}},\ \bibinfo {pages} {1226} (\bibinfo {year}
  {1995})}\BibitemShut {NoStop}%
\bibitem [{\citenamefont {Ballerini}\ \emph {et~al.}(2008)\citenamefont
  {Ballerini}, \citenamefont {Cabibbo}, \citenamefont {Candelier},
  \citenamefont {Cavagna}, \citenamefont {Cisbani}, \citenamefont {Giardina},
  \citenamefont {Lecomte}, \citenamefont {Orlandi}, \citenamefont {Parisi},
  \citenamefont {Procaccini} \emph {et~al.}}]{ballerini2008interaction}%
  \BibitemOpen
  \bibfield  {author} {\bibinfo {author} {\bibfnamefont {M.}~\bibnamefont
  {Ballerini}}, \bibinfo {author} {\bibfnamefont {N.}~\bibnamefont {Cabibbo}},
  \bibinfo {author} {\bibfnamefont {R.}~\bibnamefont {Candelier}}, \bibinfo
  {author} {\bibfnamefont {A.}~\bibnamefont {Cavagna}}, \bibinfo {author}
  {\bibfnamefont {E.}~\bibnamefont {Cisbani}}, \bibinfo {author} {\bibfnamefont
  {I.}~\bibnamefont {Giardina}}, \bibinfo {author} {\bibfnamefont
  {V.}~\bibnamefont {Lecomte}}, \bibinfo {author} {\bibfnamefont
  {A.}~\bibnamefont {Orlandi}}, \bibinfo {author} {\bibfnamefont
  {G.}~\bibnamefont {Parisi}}, \bibinfo {author} {\bibfnamefont
  {A.}~\bibnamefont {Procaccini}},  \emph {et~al.},\ }\href@noop {} {\bibfield
  {journal} {\bibinfo  {journal} {Proceedings of the national academy of
  sciences}\ }\textbf {\bibinfo {volume} {105}},\ \bibinfo {pages} {1232}
  (\bibinfo {year} {2008})}\BibitemShut {NoStop}%
\bibitem [{\citenamefont {Chat\'e}\ \emph {et~al.}(2008)\citenamefont {Chat\'e}
  \emph {et~al.}}]{Chate08}%
  \BibitemOpen
  \bibfield  {author} {\bibinfo {author} {\bibfnamefont {H.}~\bibnamefont
  {Chat\'e}} \emph {et~al.},\ }\href {\doibase 10.1103/PhysRevE.77.046113}
  {\bibfield  {journal} {\bibinfo  {journal} {Phys. Rev. E}\ }\textbf {\bibinfo
  {volume} {77}},\ \bibinfo {pages} {046113} (\bibinfo {year}
  {2008})}\BibitemShut {NoStop}%
\bibitem [{\citenamefont {Deseigne}\ \emph {et~al.}(2010)\citenamefont
  {Deseigne}, \citenamefont {Dauchot},\ and\ \citenamefont
  {Chat\'e}}]{Deseigne10}%
  \BibitemOpen
  \bibfield  {author} {\bibinfo {author} {\bibfnamefont {J.}~\bibnamefont
  {Deseigne}}, \bibinfo {author} {\bibfnamefont {O.}~\bibnamefont {Dauchot}}, \
  and\ \bibinfo {author} {\bibfnamefont {H.}~\bibnamefont {Chat\'e}},\ }\href
  {\doibase 10.1103/PhysRevLett.105.098001} {\bibfield  {journal} {\bibinfo
  {journal} {Phys. Rev. Lett.}\ }\textbf {\bibinfo {volume} {105}},\ \bibinfo
  {pages} {098001} (\bibinfo {year} {2010})}\BibitemShut {NoStop}%
\bibitem [{\citenamefont {Theurkauff}\ \emph {et~al.}(2012)\citenamefont
  {Theurkauff}, \citenamefont {Cottin-Bizonne}, \citenamefont {Palacci},
  \citenamefont {Ybert},\ and\ \citenamefont
  {Bocquet}}]{PhysRevLett.108.268303}%
  \BibitemOpen
  \bibfield  {author} {\bibinfo {author} {\bibfnamefont {I.}~\bibnamefont
  {Theurkauff}}, \bibinfo {author} {\bibfnamefont {C.}~\bibnamefont
  {Cottin-Bizonne}}, \bibinfo {author} {\bibfnamefont {J.}~\bibnamefont
  {Palacci}}, \bibinfo {author} {\bibfnamefont {C.}~\bibnamefont {Ybert}}, \
  and\ \bibinfo {author} {\bibfnamefont {L.}~\bibnamefont {Bocquet}},\ }\href
  {\doibase 10.1103/PhysRevLett.108.268303} {\bibfield  {journal} {\bibinfo
  {journal} {Phys. Rev. Lett.}\ }\textbf {\bibinfo {volume} {108}},\ \bibinfo
  {pages} {268303} (\bibinfo {year} {2012})}\BibitemShut {NoStop}%
\bibitem [{\citenamefont {Cates}\ and\ \citenamefont
  {Tailleur}(2015)}]{Cates15}%
  \BibitemOpen
  \bibfield  {author} {\bibinfo {author} {\bibfnamefont {M.~E.}\ \bibnamefont
  {Cates}}\ and\ \bibinfo {author} {\bibfnamefont {J.}~\bibnamefont
  {Tailleur}},\ }\href {\doibase 10.1146/annurev-conmatphys-031214-014710}
  {\bibfield  {journal} {\bibinfo  {journal} {Annu. Rev. Condens. Matter
  Phys.}\ }\textbf {\bibinfo {volume} {6}},\ \bibinfo {pages} {219} (\bibinfo
  {year} {2015})}\BibitemShut {NoStop}%
\bibitem [{\citenamefont {Alvarado}\ \emph {et~al.}(2013)\citenamefont
  {Alvarado} \emph {et~al.}}]{Alvarado13}%
  \BibitemOpen
  \bibfield  {author} {\bibinfo {author} {\bibfnamefont {J.}~\bibnamefont
  {Alvarado}} \emph {et~al.},\ }\href {\doibase 10.1038/NPHYS2715} {\bibfield
  {journal} {\bibinfo  {journal} {Nature Physics}\ }\textbf {\bibinfo {volume}
  {9}},\ \bibinfo {pages} {591} (\bibinfo {year} {2013})}\BibitemShut {NoStop}%
\bibitem [{\citenamefont {Prost}\ and\ \citenamefont
  {Bruinsma}(1996)}]{Prost96}%
  \BibitemOpen
  \bibfield  {author} {\bibinfo {author} {\bibfnamefont {J.}~\bibnamefont
  {Prost}}\ and\ \bibinfo {author} {\bibfnamefont {R.}~\bibnamefont
  {Bruinsma}},\ }\href {http://stacks.iop.org/0295-5075/33/i=4/a=321}
  {\bibfield  {journal} {\bibinfo  {journal} {EPL}\ }\textbf {\bibinfo {volume}
  {33}},\ \bibinfo {pages} {321} (\bibinfo {year} {1996})}\BibitemShut
  {NoStop}%
\bibitem [{\citenamefont {Ramaswamy}\ \emph {et~al.}(2000)\citenamefont
  {Ramaswamy}, \citenamefont {Toner},\ and\ \citenamefont
  {Prost}}]{Ramaswamy00}%
  \BibitemOpen
  \bibfield  {author} {\bibinfo {author} {\bibfnamefont {S.}~\bibnamefont
  {Ramaswamy}}, \bibinfo {author} {\bibfnamefont {J.}~\bibnamefont {Toner}}, \
  and\ \bibinfo {author} {\bibfnamefont {J.}~\bibnamefont {Prost}},\ }\href
  {\doibase 10.1103/PhysRevLett.84.3494} {\bibfield  {journal} {\bibinfo
  {journal} {Phys. Rev. Lett.}\ }\textbf {\bibinfo {volume} {84}},\ \bibinfo
  {pages} {3494} (\bibinfo {year} {2000})}\BibitemShut {NoStop}%
\bibitem [{\citenamefont {Chen}(2004)}]{Chen04}%
  \BibitemOpen
  \bibfield  {author} {\bibinfo {author} {\bibfnamefont {H.-Y.}\ \bibnamefont
  {Chen}},\ }\href {\doibase 10.1103/PhysRevLett.92.168101} {\bibfield
  {journal} {\bibinfo  {journal} {Phys. Rev. Lett.}\ }\textbf {\bibinfo
  {volume} {92}},\ \bibinfo {pages} {168101} (\bibinfo {year}
  {2004})}\BibitemShut {NoStop}%
\bibitem [{\citenamefont {Casimir}(1948)}]{Casimir48}%
  \BibitemOpen
  \bibfield  {author} {\bibinfo {author} {\bibfnamefont {H.~B.~G.}\
  \bibnamefont {Casimir}},\ }\href@noop {} {\bibfield  {journal} {\bibinfo
  {journal} {Proc. K. Ned. Akad. Wet.}\ }\textbf {\bibinfo {volume} {51}},\
  \bibinfo {pages} {793} (\bibinfo {year} {1948})}\BibitemShut {NoStop}%
\bibitem [{\citenamefont {Kardar}\ and\ \citenamefont
  {Golestanian}(1999)}]{Kardar99}%
  \BibitemOpen
  \bibfield  {author} {\bibinfo {author} {\bibfnamefont {M.}~\bibnamefont
  {Kardar}}\ and\ \bibinfo {author} {\bibfnamefont {R.}~\bibnamefont
  {Golestanian}},\ }\href {\doibase 10.1103/RevModPhys.71.1233} {\bibfield
  {journal} {\bibinfo  {journal} {Rev. Mod. Phys.}\ }\textbf {\bibinfo {volume}
  {71}},\ \bibinfo {pages} {1233} (\bibinfo {year} {1999})}\BibitemShut
  {NoStop}%
\bibitem [{\citenamefont {Fisher}\ and\ \citenamefont
  {de~Gennes}(1978)}]{Fisher78}%
  \BibitemOpen
  \bibfield  {author} {\bibinfo {author} {\bibfnamefont {M.~E.}\ \bibnamefont
  {Fisher}}\ and\ \bibinfo {author} {\bibfnamefont {P.~G.}\ \bibnamefont
  {de~Gennes}},\ }\href@noop {} {\bibfield  {journal} {\bibinfo  {journal} {C.
  R. Acad. Sci. Paris Ser. B}\ }\textbf {\bibinfo {volume} {287}},\ \bibinfo
  {pages} {207} (\bibinfo {year} {1978})}\BibitemShut {NoStop}%
\bibitem [{\citenamefont {Hertlein}\ \emph {et~al.}(2008)\citenamefont
  {Hertlein} \emph {et~al.}}]{Hertlein08}%
  \BibitemOpen
  \bibfield  {author} {\bibinfo {author} {\bibfnamefont {C.}~\bibnamefont
  {Hertlein}} \emph {et~al.},\ }\href {\doibase 10.1038/nature06443} {\bibfield
   {journal} {\bibinfo  {journal} {Nature}\ }\textbf {\bibinfo {volume}
  {451}},\ \bibinfo {pages} {172} (\bibinfo {year} {2008})}\BibitemShut
  {NoStop}%
\bibitem [{\citenamefont {Ajdari}\ \emph {et~al.}(1991)\citenamefont {Ajdari},
  \citenamefont {Peliti},\ and\ \citenamefont {Prost}}]{Ajdari91}%
  \BibitemOpen
  \bibfield  {author} {\bibinfo {author} {\bibfnamefont {A.}~\bibnamefont
  {Ajdari}}, \bibinfo {author} {\bibfnamefont {L.}~\bibnamefont {Peliti}}, \
  and\ \bibinfo {author} {\bibfnamefont {J.}~\bibnamefont {Prost}},\ }\href
  {\doibase 10.1103/PhysRevLett.66.1481} {\bibfield  {journal} {\bibinfo
  {journal} {Phys. Rev. Lett.}\ }\textbf {\bibinfo {volume} {66}},\ \bibinfo
  {pages} {1481} (\bibinfo {year} {1991})}\BibitemShut {NoStop}%
\bibitem [{\citenamefont {Poulin}\ \emph {et~al.}(1997)\citenamefont {Poulin}
  \emph {et~al.}}]{Poulin97}%
  \BibitemOpen
  \bibfield  {author} {\bibinfo {author} {\bibfnamefont {P.}~\bibnamefont
  {Poulin}} \emph {et~al.},\ }\href {\doibase 10.1126/science.275.5307.1770}
  {\bibfield  {journal} {\bibinfo  {journal} {Science}\ }\textbf {\bibinfo
  {volume} {275}},\ \bibinfo {pages} {1770} (\bibinfo {year}
  {1997})}\BibitemShut {NoStop}%
\bibitem [{\citenamefont {Nicolson}(1949)}]{Nicolson49}%
  \BibitemOpen
  \bibfield  {author} {\bibinfo {author} {\bibfnamefont {M.~M.}\ \bibnamefont
  {Nicolson}},\ }\href@noop {} {\bibfield  {journal} {\bibinfo  {journal}
  {Proc. Cambridge Philos. Soc.}\ }\textbf {\bibinfo {volume} {45}},\ \bibinfo
  {pages} {288} (\bibinfo {year} {1949})}\BibitemShut {NoStop}%
\bibitem [{\citenamefont {Hu}\ and\ \citenamefont {Bush}(2005)}]{Hu05Nat}%
  \BibitemOpen
  \bibfield  {author} {\bibinfo {author} {\bibfnamefont {D.~L.}\ \bibnamefont
  {Hu}}\ and\ \bibinfo {author} {\bibfnamefont {J.~W.~M.}\ \bibnamefont
  {Bush}},\ }\href {\doibase 10.1038/nature03995} {\bibfield  {journal}
  {\bibinfo  {journal} {Nature}\ }\textbf {\bibinfo {volume} {437}},\ \bibinfo
  {pages} {733} (\bibinfo {year} {2005})}\BibitemShut {NoStop}%
\bibitem [{\citenamefont {Goulian}\ \emph {et~al.}(1993)\citenamefont
  {Goulian}, \citenamefont {Bruinsma},\ and\ \citenamefont
  {Pincus}}]{Goulian93EPL}%
  \BibitemOpen
  \bibfield  {author} {\bibinfo {author} {\bibfnamefont {M.}~\bibnamefont
  {Goulian}}, \bibinfo {author} {\bibfnamefont {R.}~\bibnamefont {Bruinsma}}, \
  and\ \bibinfo {author} {\bibfnamefont {P.}~\bibnamefont {Pincus}},\ }\href
  {http://stacks.iop.org/0295-5075/22/i=2/a=012} {\bibfield  {journal}
  {\bibinfo  {journal} {EPL}\ }\textbf {\bibinfo {volume} {22}},\ \bibinfo
  {pages} {145} (\bibinfo {year} {1993})}\BibitemShut {NoStop}%
\bibitem [{\citenamefont {Dan}\ \emph {et~al.}(1993)\citenamefont {Dan},
  \citenamefont {Pincus},\ and\ \citenamefont {Safran}}]{Dan93Lang}%
  \BibitemOpen
  \bibfield  {author} {\bibinfo {author} {\bibfnamefont {N.}~\bibnamefont
  {Dan}}, \bibinfo {author} {\bibfnamefont {P.}~\bibnamefont {Pincus}}, \ and\
  \bibinfo {author} {\bibfnamefont {S.~A.}\ \bibnamefont {Safran}},\ }\href
  {\doibase 10.1021/la00035a005} {\bibfield  {journal} {\bibinfo  {journal}
  {Langmuir}\ }\textbf {\bibinfo {volume} {9}},\ \bibinfo {pages} {2768}
  (\bibinfo {year} {1993})}\BibitemShut {NoStop}%
\bibitem [{\citenamefont {Dommersnes}\ and\ \citenamefont
  {Fournier}(1999)}]{Dommersnes99EPJB}%
  \BibitemOpen
  \bibfield  {author} {\bibinfo {author} {\bibfnamefont {P.}~\bibnamefont
  {Dommersnes}}\ and\ \bibinfo {author} {\bibfnamefont {J.-B.}\ \bibnamefont
  {Fournier}},\ }\href {\doibase 10.1007/s100510050968} {\bibfield  {journal}
  {\bibinfo  {journal} {Eur. Phys. J. B}\ }\textbf {\bibinfo {volume} {12}},\
  \bibinfo {pages} {9} (\bibinfo {year} {1999})}\BibitemShut {NoStop}%
\bibitem [{\citenamefont {Bitbol}\ \emph {et~al.}(2012)\citenamefont {Bitbol},
  \citenamefont {Constantin},\ and\ \citenamefont {Fournier}}]{BitbolPlos12}%
  \BibitemOpen
  \bibfield  {author} {\bibinfo {author} {\bibfnamefont {A.-F.}\ \bibnamefont
  {Bitbol}}, \bibinfo {author} {\bibfnamefont {D.}~\bibnamefont {Constantin}},
  \ and\ \bibinfo {author} {\bibfnamefont {J.-B.}\ \bibnamefont {Fournier}},\
  }\href {\doibase 10.1371/journal.pone.0048306} {\bibfield  {journal}
  {\bibinfo  {journal} {PLoS ONE}\ }\textbf {\bibinfo {volume} {7}},\ \bibinfo
  {pages} {e48306} (\bibinfo {year} {2012})}\BibitemShut {NoStop}%
\bibitem [{\citenamefont {van~der Wel}\ \emph {et~al.}(2016)\citenamefont
  {van~der Wel}, \citenamefont {Vahid}, \citenamefont {Šarić}, \citenamefont
  {Idema}, \citenamefont {Heinrich},\ and\ \citenamefont
  {Kraft}}]{vanderWel16}%
  \BibitemOpen
  \bibfield  {author} {\bibinfo {author} {\bibfnamefont {C.}~\bibnamefont
  {van~der Wel}}, \bibinfo {author} {\bibfnamefont {A.}~\bibnamefont {Vahid}},
  \bibinfo {author} {\bibfnamefont {A.}~\bibnamefont {Šarić}}, \bibinfo
  {author} {\bibfnamefont {T.}~\bibnamefont {Idema}}, \bibinfo {author}
  {\bibfnamefont {D.}~\bibnamefont {Heinrich}}, \ and\ \bibinfo {author}
  {\bibfnamefont {D.~J.}\ \bibnamefont {Kraft}},\ }\href {\doibase
  10.1038/srep32825} {\bibfield  {journal} {\bibinfo  {journal} {Scientific
  Reports}\ }\textbf {\bibinfo {volume} {6}} (\bibinfo {year} {2016}),\
  10.1038/srep32825}\BibitemShut {NoStop}%
\bibitem [{\citenamefont {Lu}\ \emph {et~al.}(2015)\citenamefont {Lu},
  \citenamefont {Dean},\ and\ \citenamefont {Podgornik}}]{lu2015out}%
  \BibitemOpen
  \bibfield  {author} {\bibinfo {author} {\bibfnamefont {B.-S.}\ \bibnamefont
  {Lu}}, \bibinfo {author} {\bibfnamefont {D.~S.}\ \bibnamefont {Dean}}, \ and\
  \bibinfo {author} {\bibfnamefont {R.}~\bibnamefont {Podgornik}},\ }\href@noop
  {} {\bibfield  {journal} {\bibinfo  {journal} {EPL (Europhysics Letters)}\
  }\textbf {\bibinfo {volume} {112}},\ \bibinfo {pages} {20001} (\bibinfo
  {year} {2015})}\BibitemShut {NoStop}%
\bibitem [{\citenamefont {Chen}\ and\ \citenamefont {Chen}(2006)}]{Chen06}%
  \BibitemOpen
  \bibfield  {author} {\bibinfo {author} {\bibfnamefont {C.-H.}\ \bibnamefont
  {Chen}}\ and\ \bibinfo {author} {\bibfnamefont {H.-Y.}\ \bibnamefont
  {Chen}},\ }\href {\doibase 10.1103/PhysRevE.74.051917} {\bibfield  {journal}
  {\bibinfo  {journal} {Phys. Rev. E}\ }\textbf {\bibinfo {volume} {74}},\
  \bibinfo {pages} {051917} (\bibinfo {year} {2006})}\BibitemShut {NoStop}%
\bibitem [{\citenamefont {Chen}\ and\ \citenamefont
  {Mikhailov}(2010)}]{Chen10}%
  \BibitemOpen
  \bibfield  {author} {\bibinfo {author} {\bibfnamefont {H.-Y.}\ \bibnamefont
  {Chen}}\ and\ \bibinfo {author} {\bibfnamefont {A.~S.}\ \bibnamefont
  {Mikhailov}},\ }\href {\doibase 10.1103/PhysRevE.81.031901} {\bibfield
  {journal} {\bibinfo  {journal} {Phys. Rev. E}\ }\textbf {\bibinfo {volume}
  {81}},\ \bibinfo {pages} {031901} (\bibinfo {year} {2010})}\BibitemShut
  {NoStop}%
\bibitem [{\citenamefont {Oono}\ and\ \citenamefont {Shiwa}(1987)}]{Oono87}%
  \BibitemOpen
  \bibfield  {author} {\bibinfo {author} {\bibfnamefont {Y.}~\bibnamefont
  {Oono}}\ and\ \bibinfo {author} {\bibfnamefont {Y.}~\bibnamefont {Shiwa}},\
  }\href {\doibase 10.1142/S0217984987000077} {\bibfield  {journal} {\bibinfo
  {journal} {Modern Physics Letters B}\ }\textbf {\bibinfo {volume} {01}},\
  \bibinfo {pages} {49} (\bibinfo {year} {1987})}\BibitemShut {NoStop}%
\bibitem [{\citenamefont {Oono}\ and\ \citenamefont {Bahiana}(1988)}]{Oono88}%
  \BibitemOpen
  \bibfield  {author} {\bibinfo {author} {\bibfnamefont {Y.}~\bibnamefont
  {Oono}}\ and\ \bibinfo {author} {\bibfnamefont {M.}~\bibnamefont {Bahiana}},\
  }\href {https://journals.aps.org/prl/abstract/10.1103/PhysRevLett.61.1109}
  {\bibfield  {journal} {\bibinfo  {journal} {Physical review letters}\
  }\textbf {\bibinfo {volume} {61}},\ \bibinfo {pages} {1109} (\bibinfo {year}
  {1988})}\BibitemShut {NoStop}%
\bibitem [{\citenamefont {Puri}\ and\ \citenamefont {Frisch}(1998)}]{Puri98}%
  \BibitemOpen
  \bibfield  {author} {\bibinfo {author} {\bibfnamefont {S.}~\bibnamefont
  {Puri}}\ and\ \bibinfo {author} {\bibfnamefont {H.~L.}\ \bibnamefont
  {Frisch}},\ }\href {\doibase 10.1142/S0217979298000892} {\bibfield  {journal}
  {\bibinfo  {journal} {International Journal of Modern Physics B}\ }\textbf
  {\bibinfo {volume} {12}},\ \bibinfo {pages} {1623} (\bibinfo {year}
  {1998})}\BibitemShut {NoStop}%
\bibitem [{\citenamefont {Glotzer}\ \emph {et~al.}(1994)\citenamefont
  {Glotzer}, \citenamefont {Stauffer},\ and\ \citenamefont {Jan}}]{Glotzer94}%
  \BibitemOpen
  \bibfield  {author} {\bibinfo {author} {\bibfnamefont {S.~C.}\ \bibnamefont
  {Glotzer}}, \bibinfo {author} {\bibfnamefont {D.}~\bibnamefont {Stauffer}}, \
  and\ \bibinfo {author} {\bibfnamefont {N.}~\bibnamefont {Jan}},\ }\href
  {https://journals.aps.org/prl/abstract/10.1103/PhysRevLett.72.4109}
  {\bibfield  {journal} {\bibinfo  {journal} {Physical review letters}\
  }\textbf {\bibinfo {volume} {72}},\ \bibinfo {pages} {4109} (\bibinfo {year}
  {1994})}\BibitemShut {NoStop}%
\bibitem [{\citenamefont {Glotzer}\ \emph {et~al.}(1995)\citenamefont
  {Glotzer}, \citenamefont {Di~Marzio},\ and\ \citenamefont
  {Muthukumar}}]{Glotzer95}%
  \BibitemOpen
  \bibfield  {author} {\bibinfo {author} {\bibfnamefont {S.~C.}\ \bibnamefont
  {Glotzer}}, \bibinfo {author} {\bibfnamefont {E.~A.}\ \bibnamefont
  {Di~Marzio}}, \ and\ \bibinfo {author} {\bibfnamefont {M.}~\bibnamefont
  {Muthukumar}},\ }\href
  {https://journals.aps.org/prl/abstract/10.1103/PhysRevLett.74.2034}
  {\bibfield  {journal} {\bibinfo  {journal} {Physical review letters}\
  }\textbf {\bibinfo {volume} {74}},\ \bibinfo {pages} {2034} (\bibinfo {year}
  {1995})}\BibitemShut {NoStop}%
\bibitem [{\citenamefont {Krishnan}\ and\ \citenamefont
  {Puri}(2015)}]{Krishnan15}%
  \BibitemOpen
  \bibfield  {author} {\bibinfo {author} {\bibfnamefont {R.}~\bibnamefont
  {Krishnan}}\ and\ \bibinfo {author} {\bibfnamefont {S.}~\bibnamefont
  {Puri}},\ }\href {\doibase 10.1103/PhysRevE.92.052316} {\bibfield  {journal}
  {\bibinfo  {journal} {Physical Review E}\ }\textbf {\bibinfo {volume} {92}}
  (\bibinfo {year} {2015}),\ 10.1103/PhysRevE.92.052316}\BibitemShut {NoStop}%
\bibitem [{\citenamefont {Komura}\ and\ \citenamefont
  {Andelman}(2014)}]{KOMURA201434}%
  \BibitemOpen
  \bibfield  {author} {\bibinfo {author} {\bibfnamefont {S.}~\bibnamefont
  {Komura}}\ and\ \bibinfo {author} {\bibfnamefont {D.}~\bibnamefont
  {Andelman}},\ }\href {\doibase https://doi.org/10.1016/j.cis.2013.12.003}
  {\bibfield  {journal} {\bibinfo  {journal} {Advances in Colloid and Interface
  Science}\ }\textbf {\bibinfo {volume} {208}},\ \bibinfo {pages} {34 }
  (\bibinfo {year} {2014})},\ \bibinfo {note} {special issue in honour of
  Wolfgang Helfrich}\BibitemShut {NoStop}%
\bibitem [{\citenamefont {Bu}\ and\ \citenamefont
  {Callaway}(2011)}]{allostery}%
  \BibitemOpen
  \bibfield  {author} {\bibinfo {author} {\bibfnamefont {Z.}~\bibnamefont
  {Bu}}\ and\ \bibinfo {author} {\bibfnamefont {D.}~\bibnamefont {Callaway}},\
  }\href {\doibase 10.1016/B978-0-12-381262-9.00005-7} {\bibfield  {journal}
  {\bibinfo  {journal} {Adv. Protein Chem. Struct. Biol.}\ }\textbf {\bibinfo
  {volume} {83}},\ \bibinfo {pages} {163} (\bibinfo {year} {2011})}\BibitemShut
  {NoStop}%
\bibitem [{\citenamefont {Krauth}(2006)}]{krauth2006statistical}%
  \BibitemOpen
  \bibfield  {author} {\bibinfo {author} {\bibfnamefont {W.}~\bibnamefont
  {Krauth}},\ }\href@noop {} {\emph {\bibinfo {title} {Statistical mechanics:
  algorithms and computations}}},\ Vol.~\bibinfo {volume} {13}\ (\bibinfo
  {publisher} {OUP Oxford},\ \bibinfo {year} {2006})\BibitemShut {NoStop}%
\bibitem [{\citenamefont {Dean}(1996)}]{dean1996langevin}%
  \BibitemOpen
  \bibfield  {author} {\bibinfo {author} {\bibfnamefont {D.~S.}\ \bibnamefont
  {Dean}},\ }\href@noop {} {\bibfield  {journal} {\bibinfo  {journal} {Journal
  of Physics A: Mathematical and General}\ }\textbf {\bibinfo {volume} {29}},\
  \bibinfo {pages} {L613} (\bibinfo {year} {1996})}\BibitemShut {NoStop}%
\bibitem [{\citenamefont {Kawasaki}(1994)}]{KAWASAKI199435}%
  \BibitemOpen
  \bibfield  {author} {\bibinfo {author} {\bibfnamefont {K.}~\bibnamefont
  {Kawasaki}},\ }\href {\doibase https://doi.org/10.1016/0378-4371(94)90533-9}
  {\bibfield  {journal} {\bibinfo  {journal} {Physica A: Statistical Mechanics
  and its Applications}\ }\textbf {\bibinfo {volume} {208}},\ \bibinfo {pages}
  {35 } (\bibinfo {year} {1994})}\BibitemShut {NoStop}%
\end{thebibliography}

\end{document}

\title{\bf{Supplemental Material:\\Field-Embedded Particles Driven by Active Flips}}
\author{Ruben Zakine, Jean-Baptiste Fournier and Fr\'ed\'eric van Wijland}
\date{\today}
\maketitle
%%%%%%%%%%%%%%%%%%%%%%%%%%%%%%%%%%%%%%%%%%%%%%

\section{Normalization}

In order to reduce the number of free parameters, we normalize lengths by the lattice spacing $a$, energies by $T$, times by $a^2/(\Gamma c)$ and we absorb $c$ in a redefinition of the field $\phi$. We thus replace $\bm x/a\to \bm x$, $\Gamma ct/a^2\to t$, {$c\phi^2/T\to\phi^2$}, $a^2r/c\to r$, $B/c\to B$, $T\mu/(\Gamma c)\to \mu$, $a^2\alpha/(\Gamma c)\to\alpha$ and $a^2\gamma/(\Gamma c)\to\gamma$.

\section{Simulation details}
\label{simulations}

We simulate our system on a two dimensional square lattice of size $L_x\times L_y$ with periodic boundary conditions. We use the dimensionless formulation given in the article. The Gaussian field is defined on each site $(i,j)$ and takes continuous real values $\phi_{ij}$. The particles move on the lattice sites and we denote $\bm r_k=(i_k,j_k)$ their position. The discretized Hamiltonian is given by
\begin{align}
H=\sum_{i,j}\left[\frac r 2 \phi_{ij}^2+\frac 1 2 (\nabla \phi_{ij})^2\right]+\sum_{k=1}^N \frac B 2 \left(\phi_k-\phi_0 S_k\right)^2,
\end{align}
where $\phi_k\equiv \phi_{i_k j_k}$ and $\nabla \phi_{ij}\equiv \left(\phi_{i+1,j}-\phi_{i,j},\phi_{i,j+1}-\phi_{i,j}\right)^T$. At each step $\Delta t$ we first choose randomly if we begin by updating the field and then the particles, or the opposite. 
\begin{itemize}
\item For the field update, each lattice site is updated according to
\begin{equation}
\phi_{ij}(t+\Delta t)=\phi_{ij}(t)+\Delta t\left(\nabla^2 \phi_{ij}(t)-r\phi_{ij}(t)-B\sum_{k=1}^N [\phi_{ij}(t)-\phi_0 S_k(t)]\delta_{i,i_k}\delta_{j,j_k}\right)+G_{ij}(0,2\Delta t),
\end{equation}
where $\nabla^2\phi_{ij}\equiv \phi_{i+1,j}+\phi_{i-1,j}+\phi_{i,j+1}+\phi_{i,j-1}-4\,\phi_{i,j}$ is the discrete Laplacian and the $G_{ij}(0,2\Delta t)$ are independent random Gaussian variables with mean $0$ and variance $2 \Delta t$.

\item For the particles' update, we choose $N$ times at random a particle among the $N$ particles and decide if it hops to a neighbouring site, flips its spin or does not move. We compute the probability of each event and we apply a tower sampling algorithm~\cite{krauth2006statistical}. We define $P_u$, $P_r$, $P_d$, $P_\ell$, $P_f$ the probabilities for a particle to move up, right, down, left, or to flip, respectively. The total energy variation when the particle $k$ moves from site $(i_k,j_k)$ to $(i'_k,j'_k)$ is given by 
\begin{align}
\Delta H_{\bm r_k\to \bm r'_k}=\frac{B}{2}(\phi_{i'_k,j'_k}-\phi_{i_k,j_k})(\phi_{i'_k,j'_k}+\phi_{i_k,j_k}-2\phi_0 S_k).
\end{align}
Similarly, the energy variation of the system when the particle $k$ flips spin at site $(i_k,j_k)$ is given by
\begin{align}
\Delta H_{S_k\to -S_k}=2 B\phi_0\phi_k S_k.
\end{align}
Hence, according to the Langevin dynamics, we take
\begin{align}
&P_u(k)=\mu \Delta t \,\exp\left[-\frac B 4 (\phi_{i_k,j_k+1}-\phi_{i_k,j_k})(\phi_{i_k,j_k+1}+\phi_{i_k,j_k}-2\phi_0 S_k)\right],\\
&P_r(k)=\mu \Delta t \,\exp\left[-\frac B 4 (\phi_{i_k+1,j_k}-\phi_{i_k,j_k})(\phi_{i_k+1,j_k}+\phi_{i_k,j_k}-2\phi_0 S_k)\right],\\
&P_d(k)=\mu \Delta t \,\exp\left[-\frac B 4 (\phi_{i_k,j_k-1}-\phi_{i_k,j_k})(\phi_{i_k,j_k-1}+\phi_{i_k,j_k}-2\phi_0 S_k)\right],\\
&P_\ell(k)=\mu \Delta t \,\exp\left[-\frac B 4 (\phi_{i_k-1,j_k}-\phi_{i_k,j_k})(\phi_{i_k-1,j_k}+\phi_{i_k,j_k}-2\phi_0 S_k)\right].
\end{align}
The dynamics of the flips depends on the type of particle considered. For SFIPs, the flipping probability is
\begin{align}
P_f(k)=\epsilon\, \Delta t\, \exp\left[-B\phi_0\phi_k S_k\right], 
\end{align}
and satisfies detailed balance, whereas for ASFIPs the flipping probability is independent of the field and is given by 
\begin{align}
P_f(k)=\begin{cases}
\alpha \Delta t &\text{if $S_k=-1$},\\
\gamma \Delta t &\text{if $S_k=+1$}.
\end{cases}
\end{align}
The flipping rate $\epsilon$ plays no role on the phase diagram in equilibrium. We take $\Delta t$ small enough to ensure that the probabilities verify
\begin{align}
P_u(k)+P_r(k)+P_d(k)+P_\ell(k)+P_f(k)<1,
\end{align}
then  the probability $P_n(k)$ that particle $k$ neither jumps nor flips is given by $P_n(k)=1-[P_u(k)+P_r(k)+P_d(k)+P_\ell(k)+P_f(k)]$.
\label{numerical simulation}
\end{itemize}

\section{Force between two particles}

In order to measure the mediated force between two particles, we perform a simulation with only two up-spin particles in the Gaussian field. The first particle is fixed at $\bm r_1=0$, and the second particle is trapped in a quadratic potential centered on site $ \bm R_0$. Hence, the Langevin Eq.~(5) for the second particle, with $\bm r_2=\bm R$, writes in dimensionless form:
\begin{align}
\frac{d \bm R}{dt}&=-\mu\frac{\partial H}{\partial \bm R}-\mu \frac{\partial H_\mathrm{quad}}{\partial \bm R}+\sqrt{2 \mu }\bm \eta_2(t),
\label{langevin}
\end{align}
with $H_\mathrm{quad}=\lambda(\bm R(t)-\bm R_0)^2/2$ the quadratic potential. In equilibrium, when the system reaches a stationary regime, taking the average of Eq.~\eqref{langevin} yields the average force
\begin{align}
&\langle\bm f\rangle=\langle -\frac{\partial H}{\partial \bm R}\rangle=\lambda (\langle \bm R\rangle-\bm R_0),
\end{align}
which corresponds to the field-mediated interaction when $\lambda$ is large enough to ensure small fluctuations of the position of particle 2.

The field-mediated force can be analytically calculated for two fixed particles using a Hubbard-Stratonovich transformation on the partition function: 
\begin{align}
\mathcal{Z}&=\int \mathcal{D}\phi \exp\left(-\frac 1 2\int d^2x \left[r \phi^2+(\bm \nabla \phi)^2\right]
-\frac B 2\int d^2x (\phi(\bm x)-\phi_0)^2\left[\delta(\bm x)+\delta(\bm x-\bm R)\right]\right)\\
&\propto \int dk_1 dk_2 \exp\left(-\frac 1 2 (k_1,k_2)A\,(k_1,k_2)^T+i\phi_0(k_1,k_2)(1,1)^T\right), 
\end{align}
with 
\begin{align}
A=\begin{pmatrix}
B^{-1}+G(0) & G(R)\cr
G(R) & B^{-1}+G(0)
\end{pmatrix},
\end{align}
in which 
\begin{align}
G(R)=\int \frac{d^2q}{(2\pi)^2}\, \frac{e^{i\bm q\cdot \bm R}}{r+q^2}=\frac{1}{2\pi} K_0(R \sqrt{r}).\label{eq_green_function}
\end{align}
Since $G(0)$ exhibits a logarithmic UV divergence, we regularize it by introducing a cutoff in Fourier space that takes into account the finite size of the lattice spacing (unity here).
Hence $G(0)$ reads 
\begin{align}
G(0)=\frac{1}{2\pi}\int_0^\pi dq \frac{q}{r+q^2}\simeq \frac{1}{2\pi}\ln(\pi r^{-1/2}),
\end{align}
for $r\ll1$.
The total free energy $U(R)=-\ln \mathcal Z$ is then given by $U=U_\mathrm C+U_\mathrm{el}$ with
\begin{align}
U_\mathrm C&=\frac 1 2\ln \left(1-  \frac{K_0(R\sqrt{r})^2}{Q^2}\right)\\
U_\mathrm{el}&=\frac{2\pi \phi_0^2}{K_0(R\sqrt{r})+Q},\label{eq_elastic}
\end{align}
with $Q=\ln(\pi r^{-1/2})+2\pi/B$. In dimensionful form, the energy $U_\mathrm{C}$ is proportional to temperature, whereas $U_\mathrm{el}$ does not depend on temperature. The interpretation is that the total  interaction energy splits into a Casimir-like contribution $U_{\mathrm C}$, and an elastic one $U_\mathrm{el}$.

\section{Role of multibody interactions}

We want to probe the effect of $N$-body interactions. To do so, we replace the quadratic coupling of the particles to the field with a linear one: 
\begin{align}
H_\mathrm{int}=\sum_{k=1}^N\kappa S_k\phi(\bm r_k),
\end{align}
where $\kappa$ sets the strength of the coupling. In equilibrium, we can integrate out the field which results in an effective Hamiltonian for the particles featuring only pairwise interactions. Indeed, for spin-up particles with local density $\rho$, the partition function reads
\begin{align}
\mathcal{Z}&=\int \mathcal{D}\phi \exp\left(-\frac 1 2\int d^2x \left[r \phi^2+(\bm \nabla \phi)^2\right]
-\kappa\int d^2x \,\phi(\bm x)\rho(\bm x)\right)\\
&\propto \exp\left(\frac{\kappa^2}{2} \int \rho(\bm x)G(\bm x-\bm y)\rho(\bm y)\right), 
\end{align}
where $G$ is the same correlator as Eq.~\eqref{eq_green_function}.
From there we read off the direct pairwise potential:
\begin{align}
U_\mathrm{el}^\mathrm{lin}(\bm x,\bm y)=-\kappa^2 G(\bm x-\bm y)=-\frac{\kappa^2}{2\pi}K_0(|\bm x-\bm y|\sqrt{r})
\end{align}
which remains independent of temperature in dimensionful form.

\begin{figure}
\includegraphics[width=6cm]{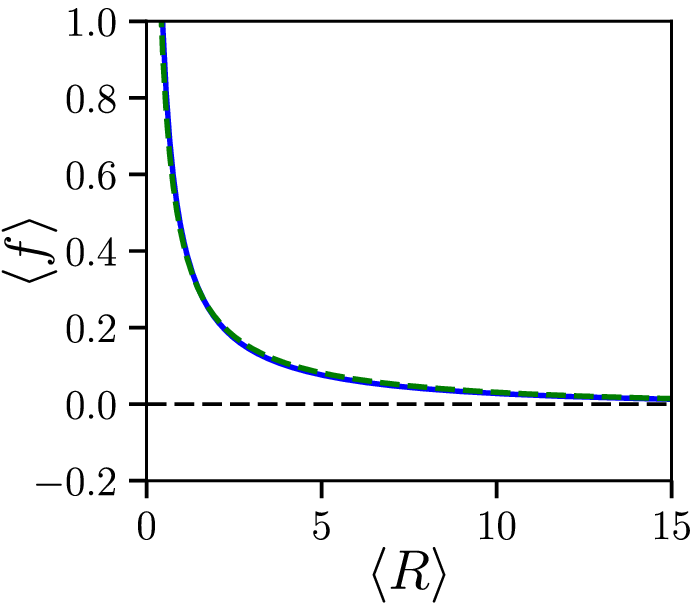}
\caption{(color online) Equilibrium average force as a function of the average particle separation. Solid blue line: analytical force deriving from the pairwise interaction when the coupling is linear $U_\mathrm{el}^\mathrm{lin}$ ($r=0.01$, $\kappa=1.7$). Dashed green line (matches exactly the blue line): analytical force deriving from the two particle interaction $U_\mathrm{el}$ when the coupling is quadratic ($r=0.01$, $B=0.26$, $\phi_0=8$).}
\label{forces}
\end{figure}

\begin{figure}
\includegraphics[width=.8\columnwidth]{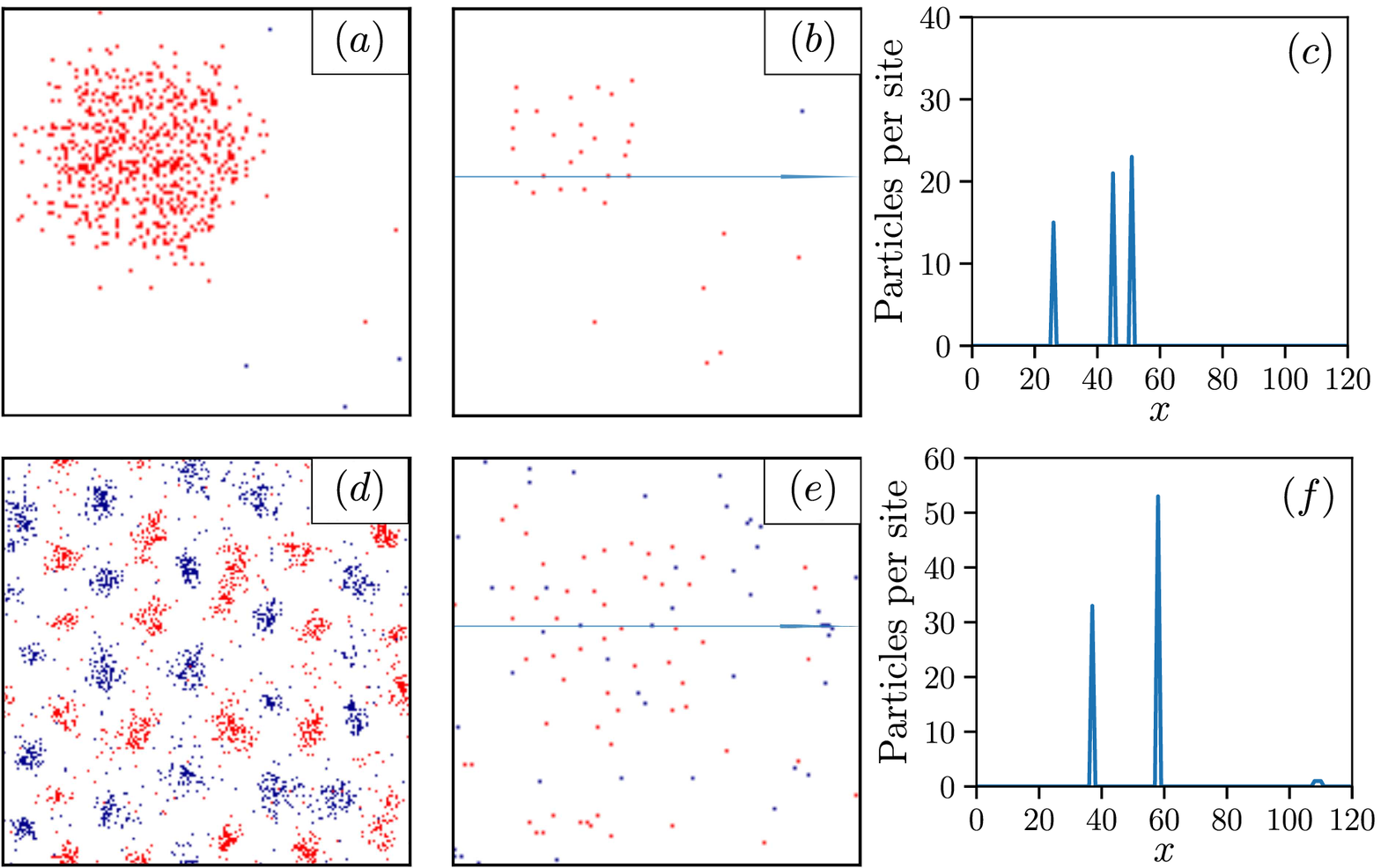}
\caption{
(a)~SFIPs quadratically coupled to the Gaussian field, with $r=0.01$, $B=0.26$, $\phi_0=8$, $\rho_0=0.05$. 
(b)~SFIPs linearly coupled to the Gaussian field with $\kappa=1.7$ tuned to yield the same interaction as the quadratic case of (a), and the same $\rho_0=0.05$, $r=0.01$.
(c)~Number of particles per site in the case of a linear coupling: 1D snapshot of (b) along the blue arrow. The system undergoes a strong and unphysical condensation. 
(d)~ASFIPs quadratically coupled to the Gaussian field, with $r=0.01$, $\rho_0=0.1$, $\mu=2.5$, $\alpha=\gamma=0.005$, $B=0.26$ and $\phi_0=8$.
(e)~ASFIPs linearly coupled to the Gaussian field with $\kappa=1.7$ tuned as before (same parameters as in (d)).
(f)~Number of particles per site in the case of a linear coupling: 1D snapshot of (e) along the blue arrow. The system undergoes a strong and unphysical condensation and patterns disappear.}
\label{fig_linear_snapshot}
\end{figure}

In order to carry out a quantitative comparison between the quadratic and the linear coupling, we adjust the parameter $\kappa$ in order to match the force obtained with the quadratic coupling. This can be done almost perfectly: as shown in Fig.~\ref{forces}, the pairwise force $F_\mathrm{lin}$ deriving from the linear coupling $U_\mathrm{el}^\mathrm{lin}$ is equivalent to the force $F_\mathrm{quad}=-\nabla U_\mathrm{el}$, with $ U_\mathrm{el}$ given by Eq.~\eqref{eq_elastic}, when $\kappa$ is correctly tuned (depending on $B$ and $\phi_0$). Then, we simulate SFIPs and ASFIPs for both couplings at matching two-body forces (fig.~\ref{fig_linear_snapshot}). In the linear case, SFIPs undergo strong unphysical condensation (see Fig.~\ref{fig_linear_snapshot}b) allowed by the absence of direct hard-core repulsion. Furthermore, patterns disappear in the ASFIP system (see Fig.~\ref{fig_linear_snapshot}e). This demonstrates the importance of multibody interactions.

\section{Movies}

\noindent Movie 1: ASFIP at the following parameters : $r=0.01$, $B=0.15$, $\phi_0=8$, $\mu=5$, $\rho_0=0.4$, and $\alpha=\gamma=0.02$.\\

\noindent Movie 2: ASFIP at the following parameters : $r=0.01$, $B=0.15$, $\phi_0=8$, $\mu=5$, $\rho_0=0.4$, and $\alpha=0.02$, $\gamma=0.0066$.

%%%%%%%%%%%%%%%%%%%%%%%%%%%%%%%%%%%%%%%%%%%%%%

%%%%%%%%%%%%%%%%%%%%%%%%%%%%%%%%%%%%%%%%%%%%%%